\documentclass[aps,pre,preprint,groupedaddress,showpacs]{revtex4}
\usepackage{graphicx}

\begin{document}

\title{Dependence of Dripping on the Orifice Diameter in a Leaky Faucet}
\author{Polina Vexler, Daniel Golubchik, Yosi Vaserman and Ariel Maniv*}
\affiliation{Physics Department, Technion - Israel Institute of
Technology, Haifa 32000,Israel}
\date{\today }

\begin{abstract}
We report the results of experiments that examined the dependence
of the dripping dynamics of a leaky faucet on the orifice
diameter. The transition of the dripping frequency between
periodic and chaotic states was found to depend on the orifice
diameter. We suggest a theoretical explanation for these
transitions based on drop formation time scales. In addition,
short-range anti-correlations were measured in the chaotic region.
These too showed a dependence on the faucet diameter. Finally, a
comparison was done between the experimental results with a
one-dimensional model for drop formation. Quantitative agreement
was found between the simulations and the experimental results.
\end{abstract}


\maketitle

\section{\label{sec:level1}Introduction}

The dripping faucet is an everyday physical system that exhibits
rich dynamical behavior. In particular it shows period doubling,
strange attractors, and transitions between numerous periodic
phases\cite{Shaw}. The leaky faucet system covers a wide range of
interest, from free surface fluid singularities (drop
formation)\cite{Eggers_3} to chaotic (dripping)
dynamics\cite{Shaw,Ros}.

Drop formation is an ancient problem\cite{Mariotte}.
Plateau\cite{Plateau} was the first to point the importance of
surface tension in generating the instability leading to drop
break-up. However a full description of drop formation was not
available until the end of the last decade, where the use of
computer models simulated the formation of single
drops\cite{Eggers_1,Eggers_2,Eggers_3,Eggers_30,Coulett,Kiyono_100,Basaran_2}.

The chaotic nature of the dripping faucet was first considered by
R\"{o}ssler\cite{Ros}. His basic prediction concerning the
behavior of a leaky faucet as a chaotic system was confirmed
experimentally by Shaw\cite{Shaw,Shaw_2}. Yet the dripping
dynamics of the leaky faucet still lacks a quantitative
theoretical description. There are two complementing theoretical
approaches that describe dripping in a leaky faucet. The first is
Shaw's mass-on-spring model\cite{Shaw} or similar
models\cite{Renna_1,Kiyono_10,Innocenzo_4,Innocenzo_5,Brito_1,Brito_2},
which give the basic characteristics of the dynamics of a leaky
faucet. However, a full quantitative description of the system can
not be deduced from such a model. The other approach is based on
solving an approximate one-dimensional form of the Navier-Stokes
equations of the
system\cite{Eggers_1,Eggers_2,Eggers_3,Eggers_30}, or a similar
one-dimensional Langrangian based fluid mechanical equations of
motion\cite{Kiyono_100,Coulett}. The use of an approximate
one-dimensional model is necessary due to the extensive
consumption of computer time using the full two-dimensional model.
However, it yields a relatively good approximate solution for drop
formation\cite{Eggers_3}, and also agrees qualitatively with the
experimental measurements of the dripping dynamics. Yet several
problems still remain using the one-dimensional model as will be
outlined below.

The lack of a quantitative description of the dripping dynamics
can be pinpointed into several basic questions. One of the main
questions is the cause for the chaotic behavior of a leaky faucet,
which is not yet clear. Shaw's model\cite{Shaw} induces an
unstable dripping state via the increasing influence of the
vibrations of the drop (as the flow rate increases) on the
dripping dynamics during drop formation. These vibrations are due
to the act of the restoring surface tension. On the other hand the
coupling between oscillations of the residue due to recoil and
drop build-up\cite{Coulett} is another interesting option.

Most of the experiments in this field focus on measuring the time
intervals between successive drops (dripping frequency). The
dripping frequency is measured as a function of the flow
rate\cite{Shaw}. Beside the obvious choice of flow rate, there are
several possible control parameters on the dripping dynamics such
as the diameter and shape of the faucet orifice, the surface
tension and the viscosity of the fluid\cite{Basaran_1}. However
the exact dependence of the dripping dynamics on these parameters
has not yet been determined. An exception is temperature. The
dependence of the dripping dynamics on temperature was measured
and quantified\cite{Katsuyama} (see below).

In the present work we focused on the dependence of the dripping
dynamics on the orifice diameter. We focused on the dependence of
the transition frequency between chaotic and periodic states on
the pipette diameter. Note that the size of the pending drop
depends on the orifice diameter. Thus the dripping frequency and
dynamics are determined also by the size of the orifice. In
addition we compared our experimental results with numerical
simulations based on the one-dimensional fluid-mechanical model
described by Fuchikami et al.\cite{Kiyono_100}.

\section{Aspects of Faucet Dynamics}

The main aspects of the leaky faucet system relevant to our
experiments are outlined below:

\subsection{Drop Formation}

Drop formation can be separated into three main
stages\cite{Coulett}:

1. Build up time, during which the drop is formed:
$\tau_{f}\sim\frac{R}{v_{0}}$, where $R$ is the orifice radius,
and $v_{0}$ is the fluid velocity.

2. Critical time (criticality), where the drop breaks-off the
fluid column: $\tau_{n}\sim\sqrt{\frac{R^{3}\rho}{\Gamma}}$, where
$\rho$ is the fluid's density, and $\Gamma$ is the surface
tension.

3. Recoil time of the residual mass after break-up:
$\tau_{d}\sim\frac{V^{\frac{7}{12}}}{\eta\sqrt{32\pi\Gamma}}$,
where $\eta$ is the viscosity, and $V$ is the residual volume.

This separation is not strict - for instance, the recoil and the
build-up stages overlap.

\subsection{Dripping Dynamics}

The dripping dynamics is influenced by many parameters. The most
relevant ones from our experimental point of view are outlined
below:

\subsubsection{\textbf{Drop Volume}}

The evolution of the drop's volume prior to break-up is
approximately linear with time. It incorporates a small
oscillatory term, at least up to dripping rates as high as $10$
drops/sec\cite{Pinto_100}. Note that the volume grows linearly
even at criticality.

Also, at the periodic state the drop volume was shown to be
approximately constant at a fixed flow rate\cite{Katsuyama}. At
this state it was claimed that the volume of the drop increased
while increasing the flow rate\cite{Katsuyama}.

\subsubsection{\textbf{Faucet Diameter}}

It has been shown \textit{qualitatively} that the dripping
dynamics depend on the diameter $d$ and thickness of the orifice
for relatively small pipettes ($d\leq4mm$) \cite{Innocenzo_3}.
Note that in our experiment all the pipettes have a diameter that
is less than $4.5mm$. The orifice wall thickness is especially
important for thin faucets in which the ratio between the orifice
wall and the inner pipette radius is less than
$0.2$\cite{Innocenzo_3}. For instance, in these thin faucets
satellite drop formation was reduced considerably compared to
thicker faucets\cite{Innocenzo_3}. However, the exact dependence
of the dripping dynamics on the orifice diameter is not yet clear.
In our experiments the ratio between the orifice wall and the
inner pipette diameter was always larger than $0.4$, thus we
expect that this factor will not influence our results. Note also
that in the following the term \emph{faucet diameter} refers to
the inner pipette diameter.

\subsubsection{\textbf{Faucet Geometry}}

The orifice geometry has a significant effect on dripping, as
shown by various
experiments\cite{Innocenzo_3,Innocenzo_1,Innocenzo_2,Pinto_5}. For
instance, increasing the asymmetry of the orifice (cut angle)
yielded a more stable dripping sequence. Increasing the faucet's
inclination softened the transition between one chaotic state to
another, from a drastic one (boundary crisis) to a smoother
one\cite{Pinto_5}. In our experiments all the faucets had a single
straight (cut angle) shape.

\subsubsection{\textbf{Temperature}}

Temperature changes influence the dripping dynamics, resulting in
an approximate linear dependence between the effective flow rate
and temperature: $dQ\sim4\times10^{-4}dT[ml/Ksec]$, for
$288K<T<303K$\cite{Katsuyama}. Note that the measured flow rate
does not change due to temperature changes, but lowering the
temperature results in a shift of the characteristic dripping
patterns to higher flow rates\cite{Whu}.

\subsubsection{\textbf{Periodic Perturbations}}

The possibility of controlling the dripping dynamics by
introducing external perturbations was examined for the leaky
faucet\cite{Shoji,Kiyono_3}. A periodic perturbation applied to
the leaky faucet changed the dripping dynamics from a stable to a
chaotic state\cite{Shoji}. Theoretical analysis of the
mass-on-spring model combined with fluid-mechanical numerical
simulations showed that a periodic perturbation induces
discontinuous transitions between chaotic and periodic dripping
states\cite{Kiyono_3}. However the exact dependence of the
dripping dynamics on external periodic perturbations is not yet
clear, and deserves further examination.

\subsection{Transitions between Chaotic and Periodic States}

\subsubsection{\textbf{Theoretical Model}}

The transition between stable (periodic) and chaotic states can be
attributed to various mechanisms\cite{Ott}. Recently there has
been an attempt to explain the origin of such transitions for
dripping dynamics\cite{Basaran_1}. The suggested model connects
between the transition from dripping to jetting to the transition
from a periodic to a chaotic state\cite{Basaran_1}. However this
model is mostly relevant to viscous fluids, which is not the case
in our experiments.

\subsubsection{\textbf{Transitions at High Flow
Rates}}

A transition from a chaotic to a stable state at high dripping
rates was one of the initial observations done for the leaky
faucet\cite{Pinto_3,Pinto_10,Pinto_11}. It was claimed that such a
transition is a result of an inverse Hopf bifurcation near
continuous flow ($f\leq40Hz$)\cite{Pinto_3,Pinto_10,Pinto_11}.

\subsection{\textbf{The Reliability of One-Dimensional
Simulations of Drop Formation}}

The comparison between theory and experiment with respect to
dripping dynamics is mostly done with the aid of one-dimensional
simulations. A comparison between the more realistic
two-dimensional (2D) Navier-Stokes equations and the approximate
one-dimensional model revealed several problems with the
approximate equations\cite{Basaran_2}. Mainly, the model can not
account for the faucet thickness, which influences drop
formation\cite{Basaran_2}, mostly for low viscous fluids under
high flow rates.

\section{Numerical Simulations}

The simulation is based on a one-dimensional Lagrangian
description of the fluid motion. It aims at reconstructing drop
formation for a fluid column stretching from an orifice. The
algorithm was first introduced by Fuchikami et
al.\cite{Kiyono_100}. Briefly, an initial drop is decomposed into
many thin disks for which separate equations of motion are written
and solved. These equations are obtained from the Lagrangian
equation under the influence of gravity, surface tension and
viscosity.

The major assumptions incorporated in the model are:
\begin{enumerate}
  \item
  The fluid is incompressible.
  \item
  The drop is axisymmetric.
  \item
  The horizontal component of the fluid velocity can be neglected in comparison with the vertical one.
  \item
  The vertical component of the velocity depends only on the vertical coordinate.
  \item
  There is no exchange of fluid between different discs.
\end{enumerate}

The $\mathcal{L}$agrangian of the system is:
\begin{equation}
    \mathcal{L}=E_{kin}-U_{g}-U_\Gamma
    \label{L}
\end{equation}

Where $E_{kin}$ is the kinetic energy, $U_{g}$ is the
gravitational energy, and $U_\Gamma$ is the surface tension, as
defined in reference no. \cite{Kiyono_100}. The resulting
$\mathcal{L}$agrangian equation of motion for the each disc is:

\begin{equation}
    \frac{d}{dt}\frac{\partial\mathcal{L}}{\partial \dot{z}_j}=\frac{\partial\mathcal{L}}{\partial z_j}+
    \frac{1}{2}\frac{\partial\dot{E}_{kin}}{\partial \dot{z}_j}
    \label{eqMotion}
\end{equation}

Where $z_j$ is the coordinate of disc $j$, and $\dot{X}$ represent
the time derivative of $X$. By integrating Eq. \ref{eqMotion} we
obtain the evolution of the drop in time and space. The
integration was done using the 5'th order Runge-Kutta
method\cite{Runge_Kutta}. After integration, the width of each
disc was checked. If the width of some disc (termed the "singular
disc") is less than a critical value (i.e. $0.01$mm), than a drop
is broken off, and the simulation restarts without the discs below
the singular disc. Thus it is possible to calculate the time
duration of drop formation from the starting point until break-off
at a specific flow rate, and therefore to compare it with the
experimental results.

\section{Experimental}

Our experimental setup consisted of a primary water tank,
connected to a laminar tube which was connected at its other end
to a pipette through which water dripped. The laminar tube was
used to stabilize the water flow. Drops falling from the pipette
interrupted a laser beam directed at a photodiode sensor. The
interruption caused a significant decrease of the sensor's
voltage, which was continuously recorded as a function of time by
a computer. The recorded data was used to calculate the time
difference between consecutive drops. The sensor's resolution was
100,000 scans per second. A schematic drawing of the experimental
setup is shown in figure \ref{fig:Setup}. In addition we used a
stroboscope with a CCD camera in order to image drop formation in
the periodic state.

\section{Experimental Results}

The time intervals between consecutive drops were measured while
continuously changing the flow rate. The flow rate was measured
directly by filling for $30$ seconds the funnel, and then weighing
it. At the same time the water level (height) in the tank was
measured. This procedure was repeated every $5$ minutes for the
entire experiment - until the tank was emptied. Poiseuille's
law\cite{Sutera} describes the flow rate through a tube under a
given pressure difference:

\begin{equation}
    Q=\frac{\Delta P\pi r^{2}}{8l\eta}
\end{equation}

Where $Q$ is the flow rate, $r$ is the tube (faucet) radius
through which the fluid flows, $\Delta P$ is the pressure
difference along the tube, and $l$ the faucet's length. Since
$Q\propto\frac{dh}{dt}$, and $\Delta P \propto h$ where $h$ is the
water level inside the tank, the flow rate is expected to decrease
exponentially. Such a decay was measured. Then after an
exponential fit based on Poiseuille's law was made to the measured
flow rate as a function of time. The result was used to calibrate
our experiments - given an initial water level it was then
possible to calculate the exact flow rate from the recorded time
that has elapsed since the beginning of the experiment. Note that
the calibration was done for each faucet separately since the flow
rate depends on the faucet diameter.

The resulting measurement of inter-drop time interval as a
function of flow rate is shown in figure \ref{fig:RawData}. Such
measurements have been done for several different faucets, each
with a specific orifice diameter. The diameters $d$ of the
pipettes ranged between $0.9mm \leq d \leq 4.4mm$.

Defining $\Delta T$ as the inter-drop time difference, it is
possible to distinguish between three major regions in figure
\ref{fig:RawData}. First a low frequency region for which $\Delta
T>120msec$ , where the dripping period is quasi-stable. This state
is roughly stable, interrupted occasionally by repeated structures
of unstable states\cite{Innocenzo_1}. A chaotic region is seen for
$50msec<\Delta T<120msec$ where there is a broad distribution of
points and occasionally a periodic window with several specific
dripping frequencies. Finally, a high frequency state for $\Delta
T<50msec$ , where a periodic state was measured for $d\geq2.3mm$
pipette diameters.

A quantitative comparison between the different dripping regions
was done by calculating the normalized average deviation from the
mean time interval between consecutive drops (SD):

\begin{equation}
SD(\overline{\Delta T})=\frac{\sqrt{\overline{\Delta
T^{2}}-\overline{\Delta T}^{2}}}{\overline{\Delta T}}
\end{equation}

where $\overline{\Delta T}$ and $\overline{\Delta T^{2}}$ are the
average values of $\Delta T$ and $\Delta T^{2}$ for 1000
consecutive drops. Note that changing the number of drops over
which the average is made does not change the resulting $SD$.
Also, over 1000 drops the maximal flow rate change is
$0.01\frac{ml}{sec}$, which is of the same order of magnitude as
the error of the measured flow rate.

The results of these calculations for an experiment using a
$4.4mm$ and a $2.3mm$ diameter pipettes are shown in figure
\ref{fig:SD}. For clarity we define a chaotic region as one in
which $SD > 0.05$.

Similar to the raw data measurements shown in figure
\ref{fig:RawData}, figure \ref{fig:SD} also shows three main
distinguishable regions: a stable area for $\Delta T\leq50msec$ ,a
chaotic area between $50msec\leq\Delta T\leq120msec$, and another
stable region for $\Delta T\geq120msec$. Note that there is a
significant difference for the transition frequency from a stable
to a chaotic state between the different faucet measurements. This
difference also exists for the transition from a chaotic to a
periodic state at high flow rates.

\subsection{Dripping Dependence on Orifice Diameter}

In order to examine the dependence of the dripping dynamics on the
orifice diameter we repeated the \emph{SD} calculations for
different pipettes. Then we extracted the flow rate at which a
transition between chaotic and periodic dripping takes place.
Figure \ref{fig:LFDep} shows the dependence of the transition
frequency on the pipette diameter $d$. As seen in figure
\ref{fig:LFDep} decreasing the orifice diameter increases the
transition frequency from a periodic to a chaotic state. As noted
above, for relatively large orifices ($d \geq 2.3mm$) there is a
transition into a high frequency stable state from a chaotic
state. As seen in the inset of figure \ref{fig:LFDep}, this
transition is also diameter dependent: increasing the orifice
diameter decreases the critical dripping frequency at which the
transition occurs. It should be noted that all the experiments
were repeated several times under different temperatures for each
faucet. Note also that the overall change in temperature between
different experiments was no more than 10K. The resulting
difference in the effective flow rate\cite{Katsuyama} is much less
than the one measured between different pipette experiments.
Therefore this change gives a lower limit on the error bars of the
measured transition frequencies. Thus, temperature and possible
external periodic noise can not account for the changes in the
transition frequencies, and it is concluded that these changes are
due to the difference in the faucet diameter.

\subsection{Anti-correlation Measurements}

Correlations between inter-drop time intervals have been measured
for the leaky faucet system\cite{Pinto_20}. These correlations
were claimed to be non-Gaussian and long-ranged\cite{Pinto_20}.

In order to quantify such correlations we define a correlation
function:

\begin{equation}
C(n)=\frac{1}{1000}\sum_{m} \{dt_{m}-\langle
dt\rangle\}\{dt_{n+m}-\langle dt\rangle\}
\end{equation}

where $dt_{n}$ is the time difference between $n$ consecutive
drops, the brackets $\langle X \rangle$ denote the average value
of $X$ (for 1000 consecutive drops), and the sum is over $m=1000$
consecutive drops. Note that in $C(n)$ small $n$($<5$)represents
short-range correlations, while large $n$($>10$) represents
long-range correlations. Also, correlations are signified by low
values while anti-correlations are emphasized by (absolute large)
negative values.

We focused on calculating short-range correlations. The
calculations were done for different faucets. A typical
calculation of short-range correlations C(1) is shown in figure
\ref{fig:MFFSim} for a $4.4mm$ diameter faucet. As shown in the
figure, near the stable/chaotic transitions ($\Delta T\sim50$msec
and $110$msec) there are strong correlations emphasized by the
relatively low value of the correlation function. However, in the
center of the chaotic area there are short-range anti-correlations
were the correlation function gains a large (absolute) negative
value. Moreover, these short-range anti-correlations also depend
on the faucet orifice. As can be clearly seen in figure
\ref{fig:MFFShift}, the dripping frequency at which $C(1)$ is
maximal increases while decreasing the orifice diameter. In order
to emphasize this point we shifted the correlation function
calculations of different faucets on a time scale $t$ equal to
$t=(2.6-d)\cdot8$, where $d$ is the faucet diameter in
millimeters. The result is shown in figure \ref{fig:Scaling},
where it can be clearly seen that all measurements collapse into a
roughly approximate single curve. However this shifting gives only
a qualitative picture since the estimated errors on the shift are
large.

\subsection{Dripping Imaging}

Several images of drop formation were taken with the aid of a
stroboscope and a CCD camera. The resulting images, done using a
$4.4mm$ pipette diameter at the periodic state, showed that the
volume of the drop is approximately constant at this state. In
addition, the images showed that the drop volume decreases while
increasing the flow rate, contrary to the claim in the
literature\cite{Katsuyama}. In order to address this question
quantitatively, we measured the average drop mass as a function of
the flow rate for different experiments. The measurements were
done by filling the funnel with drops for $30$ seconds during
which the flow rate did not change considerably. Then the water in
the funnel was weighed. At the same time using the computer record
of the photo-diode sensor, the number of drops falling from the
orifice was counted. Dividing the fluid mass by the number of
drops yielded the average drop mass. This procedure was repeated
at a period of $5$ minutes between each collection for an entire
experiment. The result of these measurements is shown in figure
\ref{fig:DropMass}. As can be clearly seen the average drop mass
decreases while increasing the flow rate.

\subsection{Numerical Simulation Results}

Numerical simulations were made for a $4.4$mm diameter faucet. The
resulting inter-drop time difference for different flow rates was
used to calculate both \emph{SD} and anti-correlations. The
\emph{SD} calculations for both simulations and measurements are
shown in figure \ref{fig:SDComp}. The comparison between
experiment and simulation is relatively good for most dripping
rates. The comparison between short range anti-correlations drawn
from simulations and measurements is shown in figure
\ref{fig:MFFSim}. Again, the agreement between theory and
experiment is relatively good. Hence the approximations made in
the numerical simulations have proven to hold also for low viscous
fluids.

\section{Discussion and Conclusions}

The dependence of the dripping dynamics on the orifice diameter
was examined. We found that the system is driven to a chaotic
state at a lower dripping frequency for larger faucets. Such a
dependence of the transition frequency on the faucet diameter can
be explained using a time-scale model for the transition between
stable and chaotic states. Specifically, we require that in the
chaotic state the ratio between the recoil time $\tau_{d}$ and the
drop build-up time $\tau_{f}$ should be greater than $1$. The
transition frequency from a stable to chaotic state is set to the
dripping frequency in which both times equalize:
$\frac{\tau_{d}}{\tau_{f}}=1$ . Note that the dripping frequency
$f$ is defined as: $f=\frac{1}{\tau_{f}+\tau_{n}}$. The transition
frequency dependence on faucet diameter stems from the fact that
the above ratio depends on it. The result is shown in figure
\ref{fig:TimeScales}. The theoretical transition frequency
decreases while increasing the faucet diameter, roughly consistent
with the experimental results that are also shown in the graph.

In addition, we found that the high frequency transition from a
chaotic to a periodic state is also faucet (diameter) dependent,
where increasing the faucet diameter decreases the transition
frequency. Again, we suggest a theoretical explanation for such a
transition. Our hypothesis is based on the assumption that when
the break-up time $\tau_{n}$ equals the drop build-up time
$\tau_{f}$, the dripping dynamics stabilize. This is due to the
fact that above this dripping frequency the important (larger)
time scale is the break-up time $\tau_{n}$, which is constant for
a given faucet. Therefore above this time scale
$\tau_{n}>\tau_{f}$ the system is in a periodic state. The results
are shown in figure \ref{fig:TimeScales}. Although in this case
the agreement between theory and experiment is only partial, it is
still significant. This is due to the fact that the tendency of
the transition frequency to decrease while increasing the faucet
diameter is the same for both experimental results and theoretical
predictions. Thus our results indicate that the transition between
chaotic and periodic states is due to the interaction between drop
formation and recoil, and possibly also to the critical state.

Finally, we measured short-time anti-correlations in the dripping
faucet experiment inside the chaotic state. These too were shown
to depend on the orifice diameter. We stress that these
measurements strengthen the connection that has been drawn above
between faucet diameter and dripping dynamics. This is emphasized
by the fact that the shape of the (short-range) correlation curve
shifts towards higher dripping frequencies while decreasing the
faucet diameter. This shift represents also a similar shift of the
transition frequencies between chaotic and periodic dripping,
probed by a totally different method.

We also compared both short-range correlations and $SD$
calculations with numerical simulations. The comparison showed
roughly good agreement between theory and experiments, emphasizing
the strength of the one-dimensional simulations in giving a
partial quantitative description of dripping dynamics.

\begin{acknowledgments}
We would like to thank the generous help of the late Moshe
Kaganovich, who inspired all of us. We would also like to thank M.
Hirshoren, A. Katz and G. Ben-Yosef for their technical support.
\end{acknowledgments}

*corresponding author: mariel@tx.technion.ac.il

\begin{figure*}[h!] 
%
%
\caption{Schematic representation of the experimental system.}
\label{fig:Setup}
\end{figure*}

\begin{figure*}[h!] 
%
%
\caption{Inter-drop time difference as a function of flow rate for
a typical experiment. The measurement was done using a $4.4mm$
diameter faucet.} \label{fig:RawData}
\end{figure*}

\begin{figure*}[h!] 
%
\caption{Typical calculation of the normalized average deviation
(SD). The solid circles represent calculations of SD for
measurements done using a $4.4mm$ diameter faucet. The hollow
squares represent the same calculations for a $2.3mm$ diameter
faucet. The dashed line at $SD=0.05$ separates the stable (below)
and chaotic (above) dripping regions.} \label{fig:SD}
\end{figure*}

\begin{figure*}[h!] 
%
%
\caption{The dependence of the transition frequency from a
periodic to a chaotic state on the orifice diameter. Note the
significant increase in the transition frequency while the orifice
diameter decreases. The inset shows the dependence of the
transition frequency from a chaotic to a periodic state (high
frequencies) on the orifice diameter.} \label{fig:LFDep}
\end{figure*}

\begin{figure*}[h!] 
%
%
\caption{The short-range correlation function C(m=1) calculated
for a $4.4mm$ diameter faucet.  The solid circles represent
calculations based on experimental measurements. The hollow
squares represent calculations based on the numerical simulations
of drop formation using the one dimensional model.}
\label{fig:MFFSim}
\end{figure*}

\begin{figure*}[h!] 
%
%
\caption{Measurements of the short-range correlation function
C(m=1) for various faucets. The measurements were shifted by a
constant value (in the y-axis) for clarity. The hollow squares
represent data for $4.4mm$ diameter faucet, the solid squares for
$2.6mm$, the hollow diamonds for $2.3mm$, the solid diamonds for
$2.0mm$, the hollow triangles for $1.9mm$, the solid triangles for
$1.4mm$, and the hollow circles for $0.9mm$ diameter faucet. Note
that no short range anti-correlations were measured for the
$0.9mm$ diameter faucet.} \label{fig:MFFShift}
\end{figure*}

\begin{figure*}[h!] 
%
%
\caption{Measurements of the short-range correlation function
C(m=1) for various faucets. The measurements are shifted linearly
as described in the text. The symbols represent measurements of
different faucets, as described in the caption of Fig. 8. Note
that all measurements merge into an approximate single curve.}
\label{fig:Scaling}
\end{figure*}

\begin{figure*}[h!] 
%
%
\caption{The dependence of the drop's mass on the flow rate. The
measurements were made for the $4.4mm$ diameter faucet.}
\label{fig:DropMass}
\end{figure*}

\begin{figure*}[h!] 
%
\caption{Comparison between the normalized average deviation (SD)
calculations based on experimental results for a 4.4mm diameter
faucet (solid squares) and for the one-dimensional model (hollow
circles).} \label{fig:SDComp}
\end{figure*}

\begin{figure*}[h!] 
%
%
\caption{Comparison between measured and estimated dripping
frequency transitions. The squares refer to a transition from a
chaotic to a periodic state, and the circles to a transition from
a stable to a chaotic state. The solid symbols represent
theoretical estimations which are explained in the text. The
hollow symbols represent measurements. Note that a transition to a
high-frequency stable state was not measured for small diameter
($d<2.0mm$) faucets. Note also that all error bars are included
inside the symbols size.} \label{fig:TimeScales}
\end{figure*}


\begin{references}

\bibitem{Shaw} R. S. Shaw, Phys. Lett. A {\bf 110,} 399 (1985).

\bibitem{Eggers_3} J. Eggers, Rev. Mod. Phys. {\bf 69,} 865 (1997).

\bibitem{Ros} O. E. R\"{o}ssler in Synergetics: A Workshop. Proceedings of the International
Workshop on Synergetics at Schloss Elmau, Bavaria, 1977, edited by
H. Haken, 174 (Springer-Verlag, New York, 1977).

\bibitem{Mariotte} E. Mariotte, Trait\`{e} du mouvement des eaux et des autres corps fluids,
E. Michallet, Paris (1686).

\bibitem{Plateau} J. Plateau, Acad. Sci. Bruxelles M\`{e}m. {\bf 23,} 5 (1849).

\bibitem{Eggers_1} J. Eggers and T.F. Dupont,J. Fluid Mech.
{\bf 262,} p. 205 (1994).

\bibitem{Eggers_2} J. Eggers, Phys. Fluids {\bf 7,} p. 941 (1995).

\bibitem{Eggers_30} J. Eggers, Phys. Rev. Lett. {\bf 71,} p. 3458 (1993).

\bibitem{Coulett} P. Coulett, L. Mahadevan, and C. Riera, Prog. Theor. Phys.
Supp., {\bf 139,} p. 507 (2000).

\bibitem{Kiyono_100} N. Fuchikami, S. Ishioka, and K. Kiyono,
J. Phys. Soc. Jap., {\bf 68,} p. 1185 (1999).

\bibitem{Basaran_2} B. Ambravaneswaran, E. D. Wilkes, and O. A.
Basaran, Phys. Fluids, {\bf 14,} p. 2606 (2002).

\bibitem{Shaw_2} \emph{The Dripping Faucet as a Model Chaotic System} by R. S. Shaw,
Aerial, Santa Cruz,California (1984).

\bibitem{Renna_1} L. Renna, Phys. Rev. E {\bf 64,} n. 046213 (2001).

\bibitem{Kiyono_10} K. Kiyono and N. Fuchikami,
J. Phys. Soc. Jap. {\bf 68,} p. 3259 (1999).

\bibitem{Innocenzo_4} A. D'Innocenzo and L. Renna,
Phys. Rev. E {\bf 58,} n. 6847 (1998).

\bibitem{Innocenzo_5} A. D'Innocenzo and L. Renna,
Phys. Rev. E {\bf 55,} n. 6776 (1997).

\bibitem{Brito_1} G. I. Sanchez-Ortiz, A. L.
Salas-Brito, Physica D {\bf 89,} p. 151 (1995).

\bibitem{Brito_2} G. I. Sanchez-Ortiz and A.L. Salas-Brito,
Phys. Lett. A {\bf 203,} p. 300 (1995).

\bibitem{Basaran_1} B. Ambravaneswaran, H. J. Subramani, S. D. Phillips, and O. A.
Basaran, Phys. Rev. Lett., {\bf 93,} n. 034501 (2004).

\bibitem{Pinto_100} M. S. F. da Rocha, J. C. Sartorelli, W. M. Goncalves,
and R. D. Pinto, Phys. Rev. E, {\bf 54,} p. 2378 (1996).

\bibitem{Katsuyama} T. Katsuyama and K. Nagata, Jour. Phys. Soc. Jap.
{\bf 68,} p. 396 (1999).

\bibitem{Innocenzo_3} A. D'Innocenzo, F. Paladini, and L. Renna,
Phys. Rev. E {\bf 65,} n. 056208 (2002).

\bibitem{Innocenzo_1} A. D'Innocenzo, F. Paladini, and L. Renna,
Phys. Rev. E {\bf 69,} n. 046204 (2004).

\bibitem{Innocenzo_2} A. D'Innocenzo, F. Paladini, and L. Renna,
Physica A {\bf 338,} p. 272 (2004).

\bibitem{Pinto_5} M.B. Reyes, R.D. Pinto, A. Tufaile, J.C. Sartorelli, Phys. Lett.
A, {\bf 300,} p. 192 (2002).

\bibitem{Whu} X. Whu and Z.A. Schelly, Physica D {\bf 40,} p. 433 (1989).

\bibitem{Shoji} M. Shoji: Nonlinear characteristics in dripping faucet,
5th Int. Workshop on Chaos/Turbulence/Complex Systems, Tsukuba,
November 6, (1998).

\bibitem{Kiyono_3} K. Kiyono and N. Fuchikami, J. Phys. Soc. J.
{\bf 71,} p. 49 (2002).

\bibitem{Ott} Chaos in Dynamical Systems, 2nd Ed. , by Edward Ott {\it %
(Cambridge University Press, 2002).}

\bibitem{Pinto_3} J.G. M. da Silva, J.C. Sartorelli, W.M. Goncalves, and
R.D. Pinto, Phys. Lett. A, {\bf 226,} p. 269 (1997).

\bibitem{Pinto_10} R.D. Pinto, J.C. Sartorelli, and W.M. Goncalves,
Physica A,  {\bf 291,} p. 244 (2001).

\bibitem{Pinto_11} R. D. Pinto , W. M. Goncalves, J. C. Sartorelli, and M.
J. de Oliveira, Phys. Rev. E, {\bf 52,} p. 6896 (1995).

\bibitem{Runge_Kutta} \emph{Numerical Recipes in C++: The Art of Scientific
Computing, 2nd edition} edited by W. H. Press and S. A. Teukolsky,
(Cambridge University Press, 2002).

\bibitem{Sutera} S.P. Sutera, Ann. Rev. F. Mech. {\bf 25,} p. 1 (1993).

\bibitem{Pinto_20} T.J. Pinto, P.M.C. de Oliveira, J.C. Sartorelli, W.M. Goncalves,
and R.D. Pinto, Phys. Rev. E, {\bf 52,} p. R2168 (1996).


\end{references}
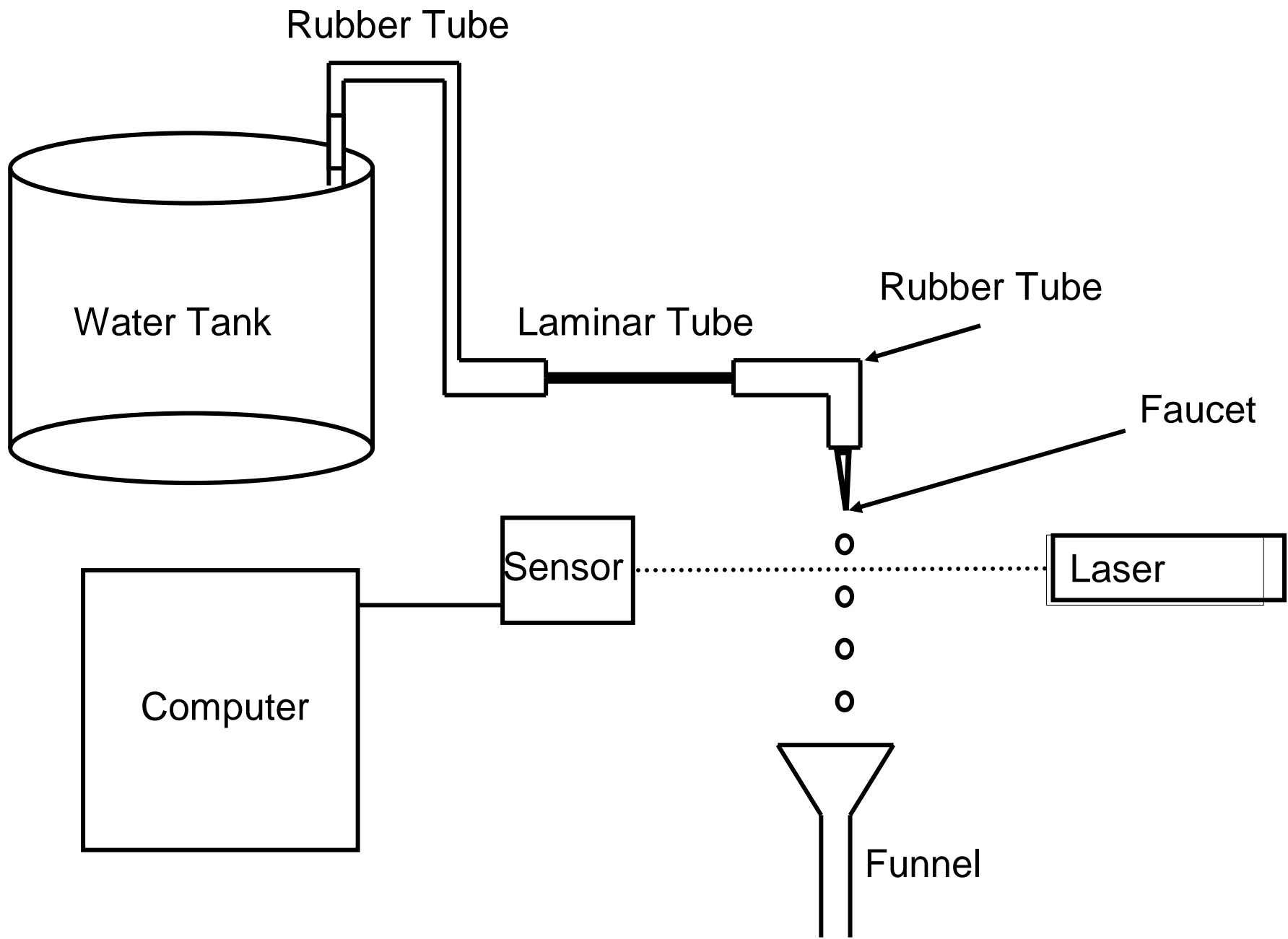
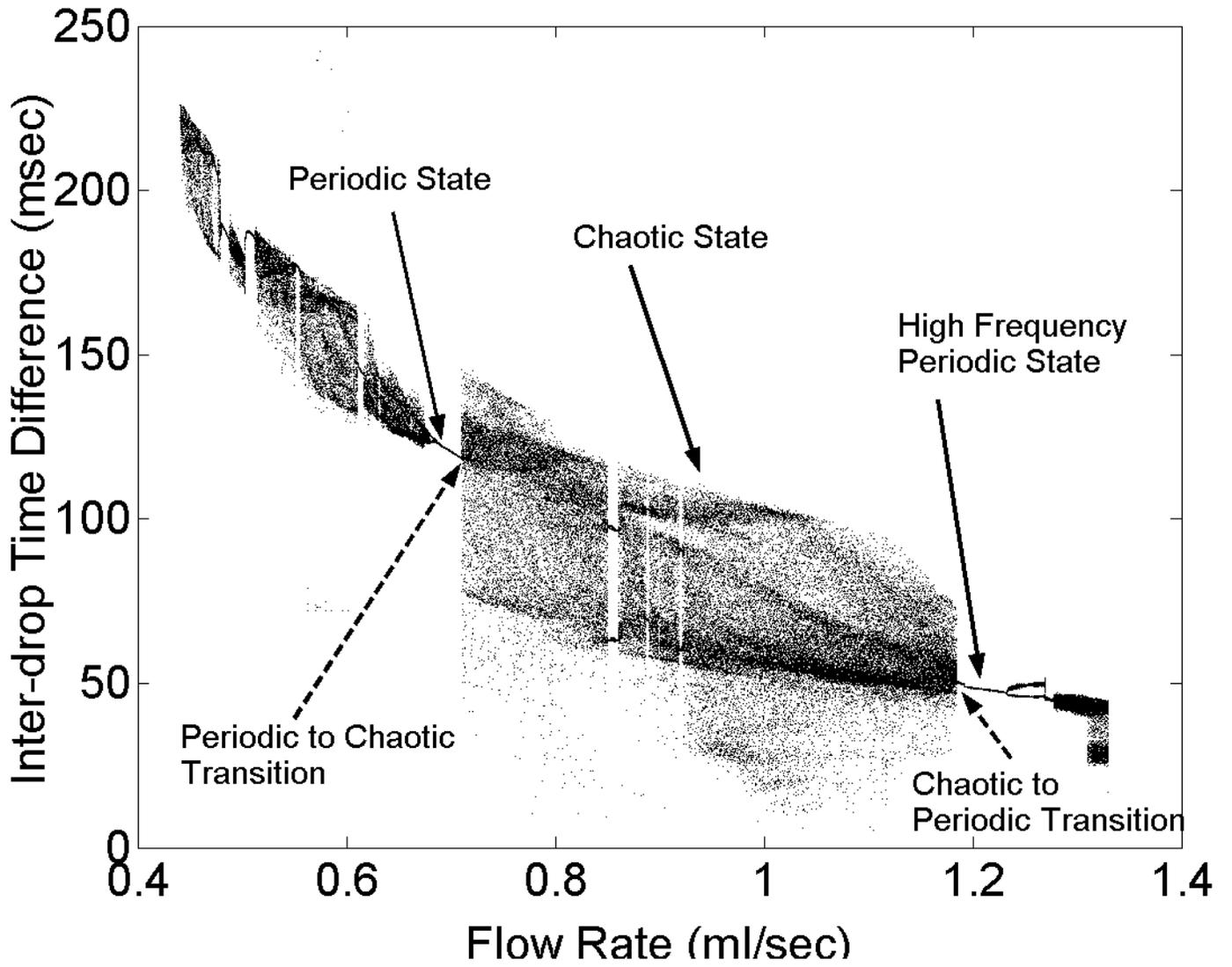
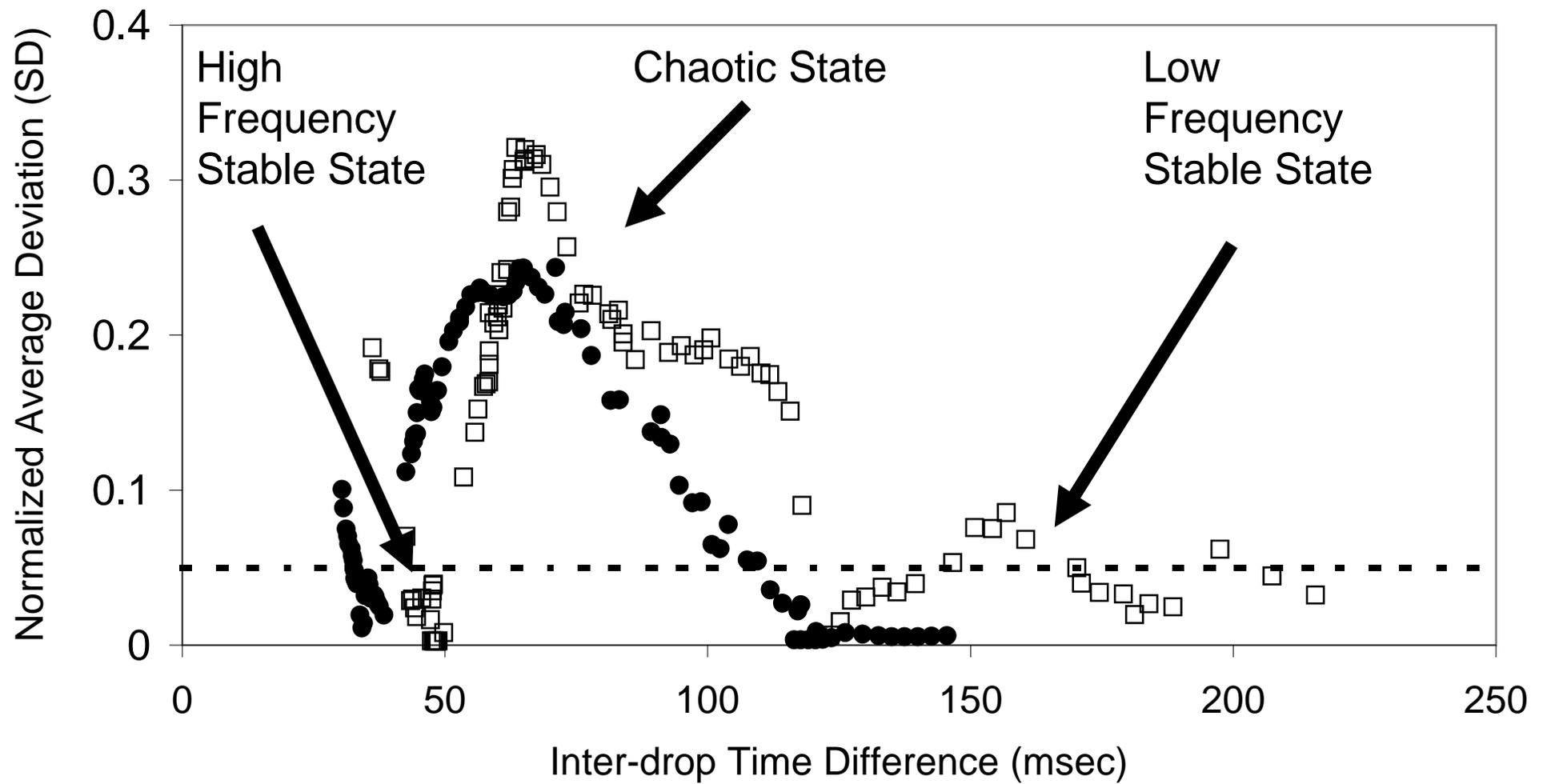
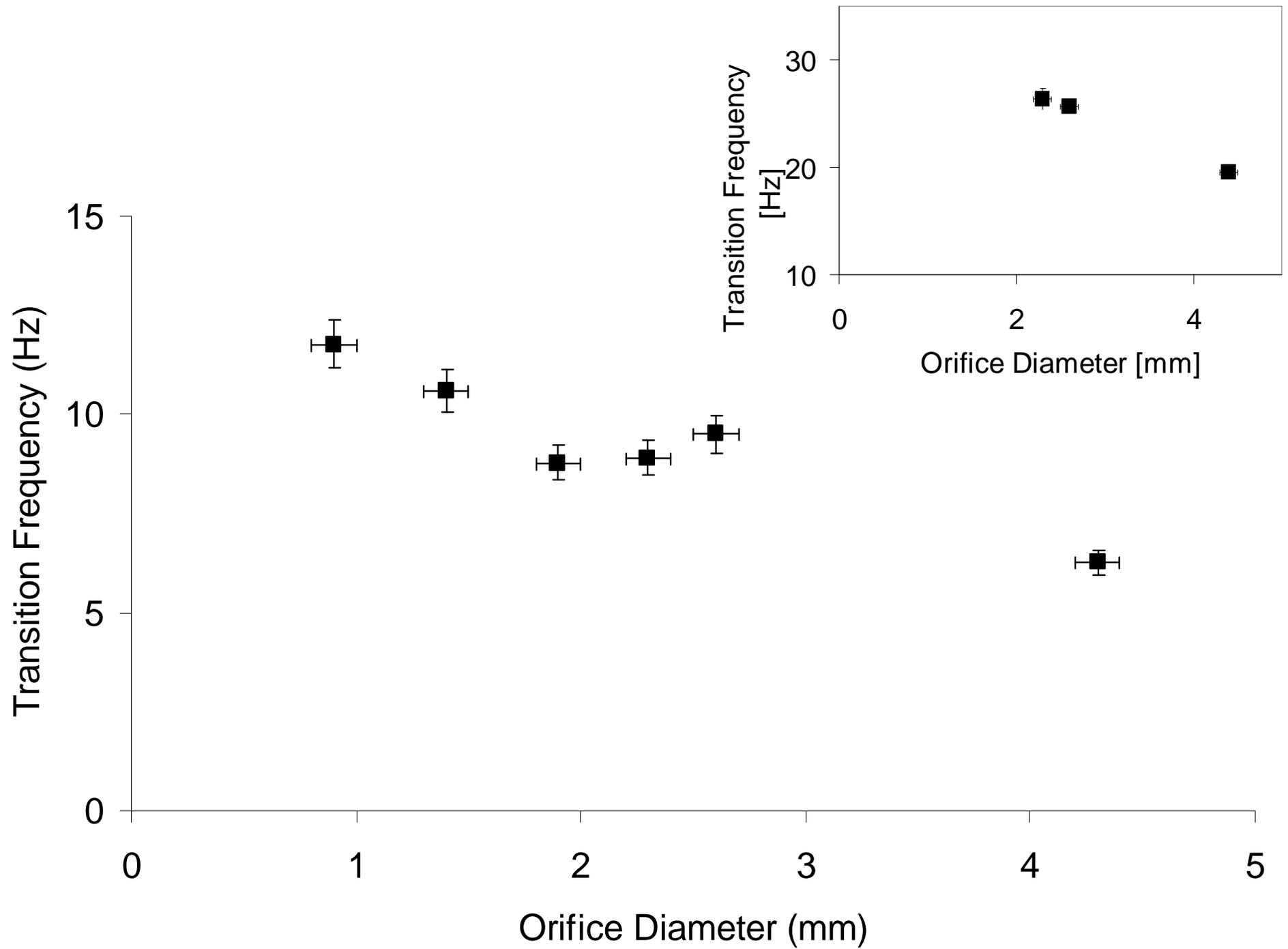
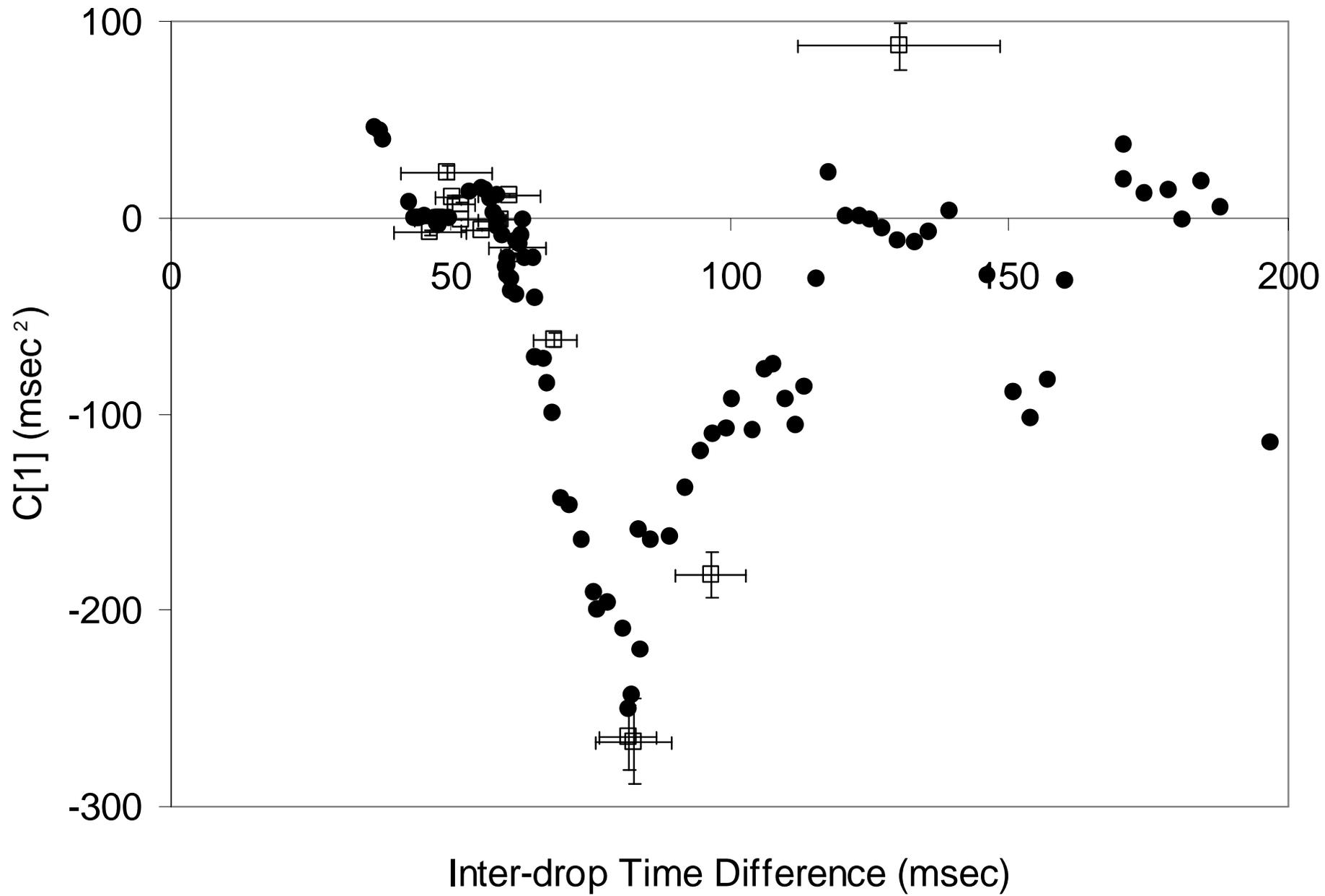
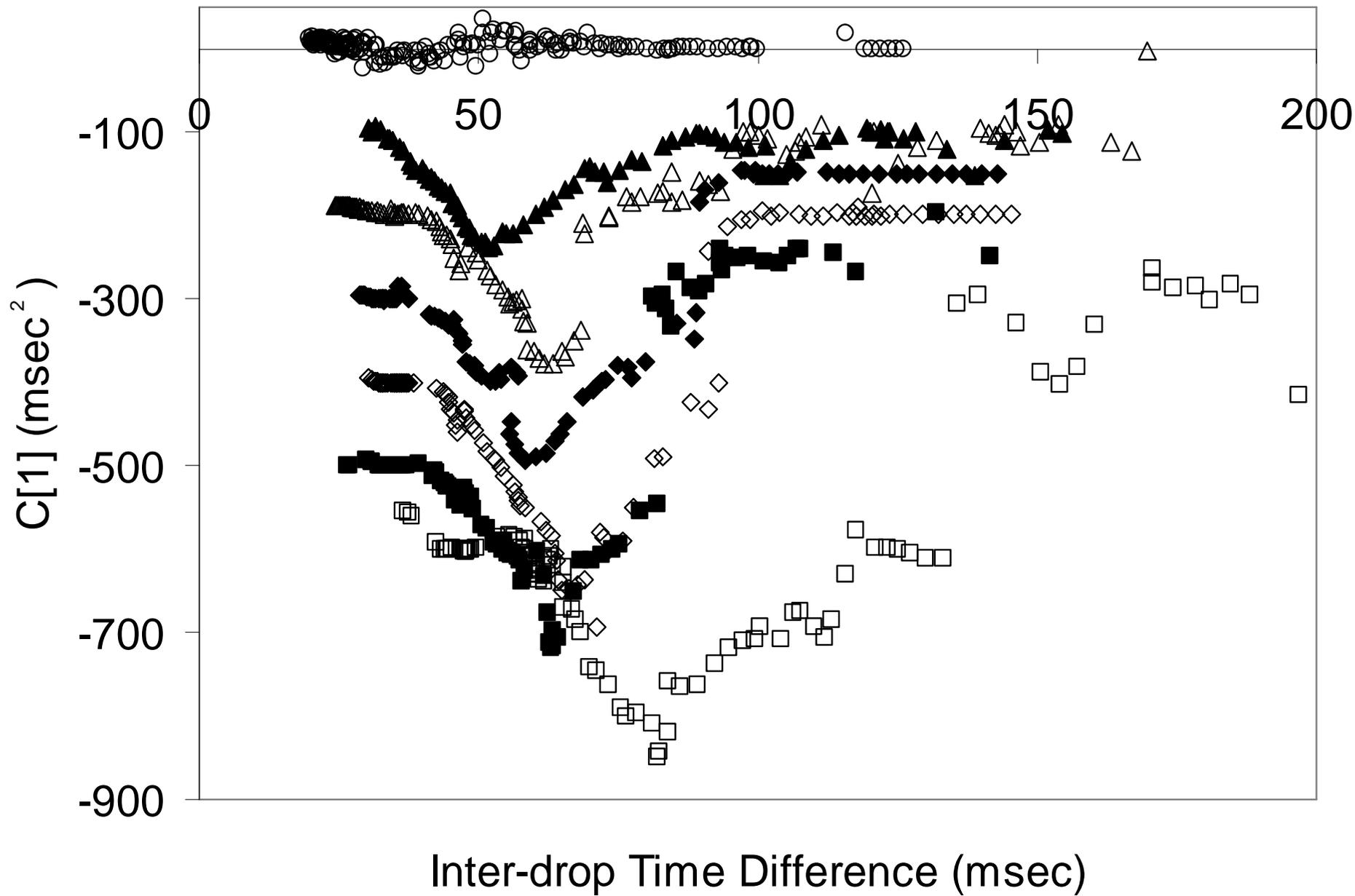
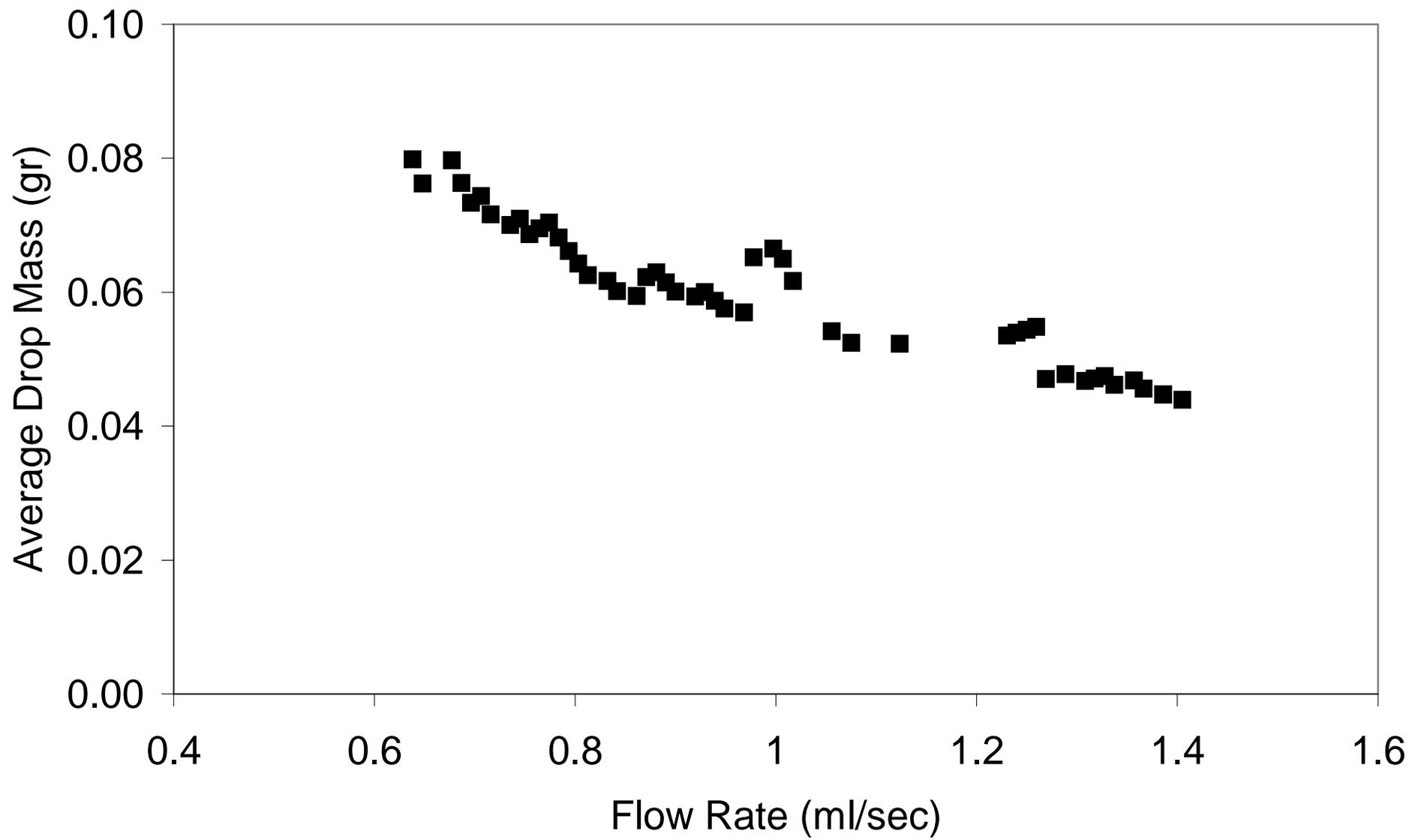
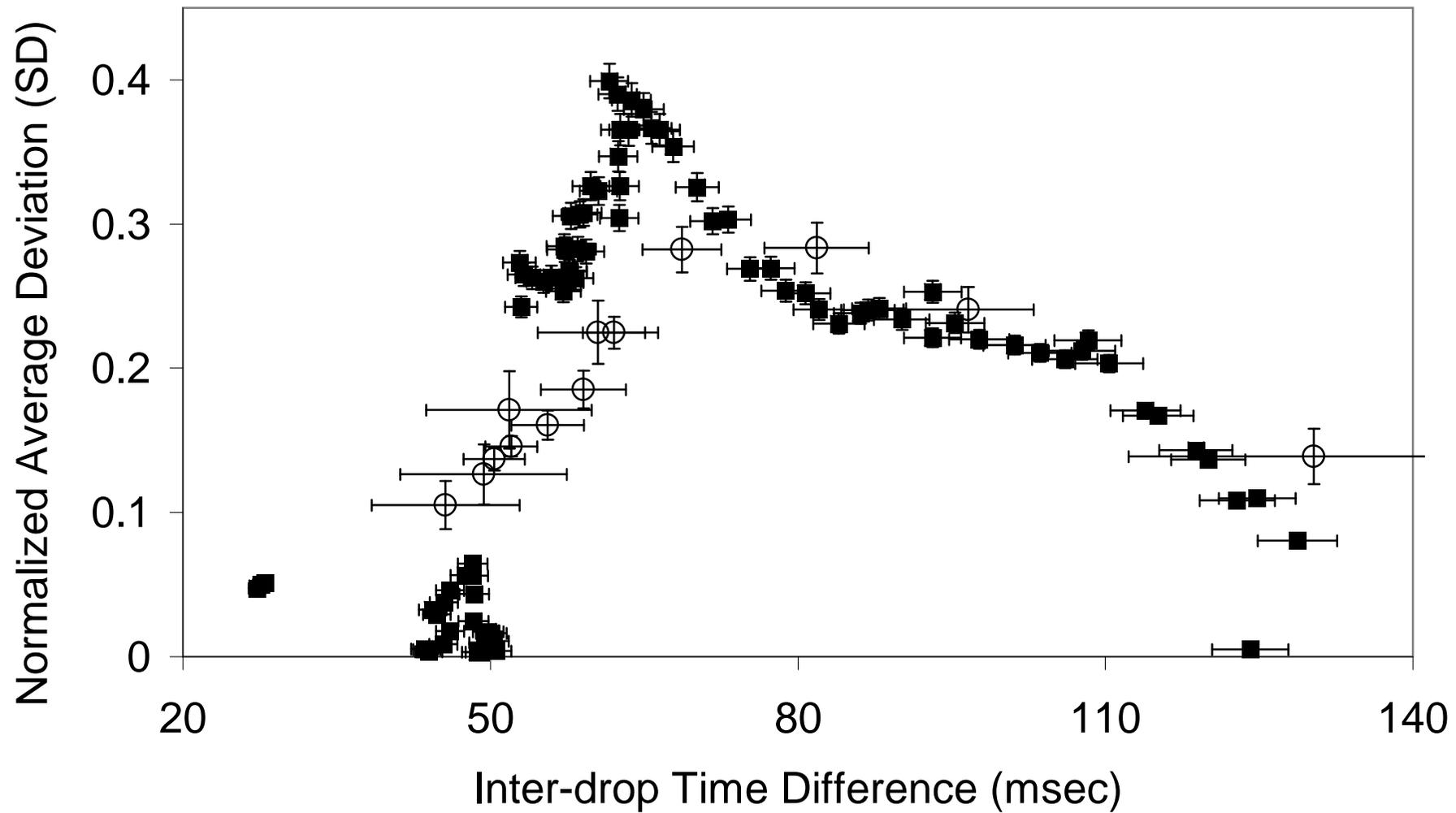
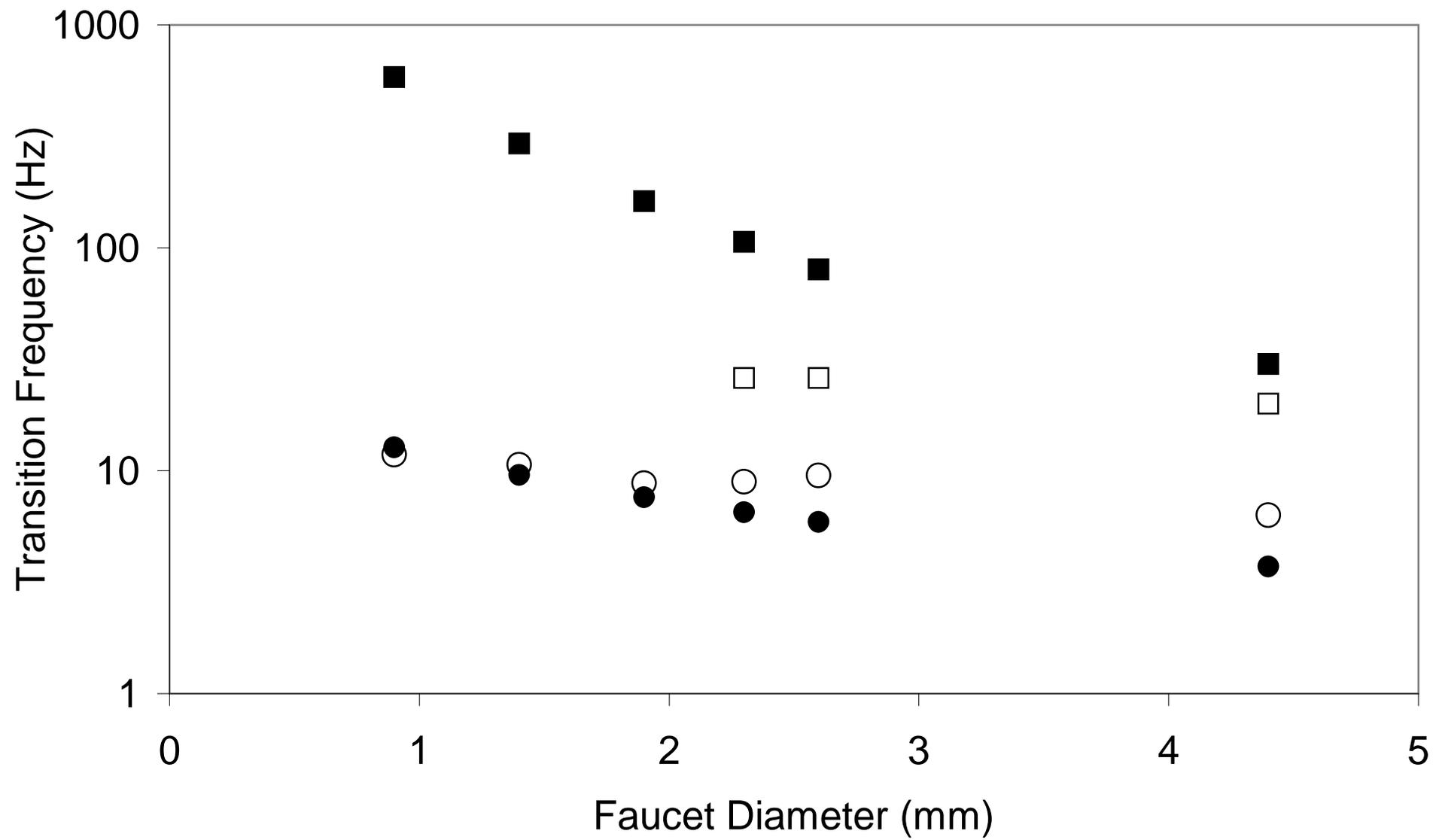
\end{document}